**Comments on "A model for fatigue in ferroelectric perovskite thin films"**

**published in Appl. Phys. Lett, 76, 1060 (2000); addendum, ibid. p.3655.**

Alexander K. Tagantsev

Ceramics Laboratiry, Materials Department, Swiss Federal Institute of Technology, EPFL

Lausanne CH-1015, Switzerland

The paper by Dawber and Scott addresses an important issue of the physics of ferroelectric thin films – the role of oxygen vacancies in polarization fatigue. However, the content of this paper [starting equations, calculations, parameters used in calculations and the way the obtained results are compared to the experimental data available in the literature] enables questioning the results and conclusions reported in it. The comments are specified in 8 issues given below.

1. To evaluate the local electric field $E_L$ at an oxygen site in the perovskite lattice the authors use an expression $E_L = E_{mac}(3\varepsilon/(2\varepsilon+1))$ [Eqs.(1a)], where $E_{mac}$ and $\varepsilon$ are the macroscopical electric field and relative dielectric constant of the material. This expression relates the macroscopical field in a spherical cavity to the external field. It basically differs from that for the local field on a crystalline site in the spherical-cavity-approximation. The latter according to the Kittel's textbook[1] reads $E_L = E_{mac} + P/(3\varepsilon_o)$ where $P$ and $\varepsilon_o$ are the macroscopic polarization in the crystal and permittivity of free space. For our problem, that implies $E_L = E_{mac}(\varepsilon+2)/3 + P_s/(3\varepsilon_o)$ where $P_s$ is the spontaneous polarization. This equation differs dramatically from that used by the authors. In the case of Pb(Zr,Ti)O$_3$ (PZT), e.g. the contributions related to $E_{mac}$ differ by a factor of about 100. In reality the difference is yet greater due to the $P_s$–term and since the local field at an oxygen site of the perovskite lattice is about 8 times larger than the result of the spherical-cavity approximation[2].

2. According to the authors the ionic transport equation reads $j = mE_L D\exp(\Delta S/k)\exp(zqbE_L/kT)$ [Eq.(2)] where $m$ is the low field defect mobility; D, defect concentration; $\Delta S$ entropy for defect movement; z, defect valence; q, electronic charge; and b, jump distance. This equation is very different from the equation $j = mD(2kT/zqb)\sinh(zqbE_{mac}/2kT)$ that follows from the basic treatment of the problem in the classical book of O'Dwyer[3] (Sec.2.1.a, Eq.(2.3)). The principal difference comes from the replacement of the macroscopical electrical field $E_{mac}$ by the local one $E_L$. For typical values of the parameters entering the formula, this substitution changes the result by two orders of magnitude at least.

3. The authors consider the oxygen vacancy redistribution as a driving force of polarization fatigue. The corresponding mathematical treatment suffers from a number of drawbacks, below I will address only the most essential one. The authors argue that the concentration of the vacancies near the electrodes exponentially grows with time when the film is periodically electrically stressed. However, as it is seen from Eq.(10), according to the developed theory, this should happen even in the absence of the driving field. On the other hand, in the absence of the driving field the problem reduces to the so-called problem of Maxwell relaxation, i.e. the evolution of a space charge density under the action of the field created by the density itself. The solution of this problem is a relaxation (exponential the linear approximation) of the charge density to a certain equilibrium value, in contrast to the exponential growth that follows form Eq.(10). Checking the authors' calculations shows that this exponential growth is a result of a wrong sign in Eq.(10). The problem with this sign is either due to a wrong sign in Eq.(6) or due to the use of a solution of the Poisson equation, which does not satisfy the boundary conditions of the problem. To finally locate the mistake one needs information on the authors' choice of the origin, which is not specified in the paper. Anyway, it is seen that the driving force of the fatigue mechanisms forwarded in the paper – i.e. exponentially growing solution of Eq.(10) - is actually the exponent of the Maxwell relaxation of fluctuations of a space charge density where one puts plus instead of minus. After this remark of mine the authors published an addendum with another derivation of the same result. As will be shown below the calculations presented in the addendum are not sound either.

4. The calculations from the addendum are based on two approximations (not indicated by the authors): (i) the diffusion contribution to the current is neglected (ii) the field of the space charge of the redistributed vacancies is not taken in to account. The first approximation might be justified however it is not clear whether it can be done for the case treated by the authors. The second approximation is inappropriate for the problem. The point is that, in electrodynamics, the field of a space charge always opposes the field responsible for the creation of this charge. For the problem in question, that implies a reduction of the field $E$ in Eq.(2) [of the addendum] with growing concentration of the oxygen vacancies. This reduction will suppress the exponential growth of the concentration, the growth that could be obtained in the approximation adopted by the authors. This can be shown on the lines of the consideration given in the main part of the paper. Thus, it turns out that the exponential growth predicted by the authors is an artifact of the used approximation. It is useful to mention that the logic of the addendum can be readily applied to the case of two non-ferroelectric back-to-back Schottky barriers resulting in a similar conclusion: an exponential growth of the carrier concentration near the electrodes. This result contradicts the present knowledge in physics of semiconductors. This also attests to the inapplicability of the approximation used by the authors.

5. According to the authors the oxygen vacancy concentration $D \propto \exp(-\Delta U/kT)$ [4] where $\Delta U = 0.7$ eV is "oxygen vacancy trapping below the Fermi level" taken from Ref.[5]. However, inspection of this paper shows that it does not deal at all with oxygen vacancies. Actually, this paper gives a value of $\Delta U = 0.7$ eV for the activation energy for the temperature dependence of the electron (hole) diffusion coefficient in a barium titanate single crystal. The paper of Dawber and Scott does not explain why the activation energy for the electron (hole) mobility in a barium titanate can be used as the activation energy for the oxygen vacancy concentration in PZT.

6. According to the authors the onset of Fowler-Nordheim tunneling occurs at applied fields an order of magnitude less that the theoretical calculations based upon the assumption of a uniform voltage drop. This contradicts the paper[6] further cited by the authors in the text, which shows that the onset of tunneling does correspond to the aforementioned calculations.

7. According to the authors their results match the temperature dependence acquired by Mihara[7]. In fact, Eqs.(19) and (20) predict a trend opposite to that shown in Fig.12 of the cited paper.

8. According to the authors Eq. (20) gives a qualitative description for the frequency dependence of the ferroelectric fatigue reported in the paper by Colla et al.[8], the fit being shown by two curve in Fig.2. Inspection of this paper, however, reveals that the comparison of the frequency dependence given by Eq. (20) to the experiment is incorrect. The point is that the difference between the two curves fitted by the authors [taken from Fig.2 from Ref.[8]] is not only the frequency but also the shape of driving pulse. Moreover, the key result of this paper is that the later factor plays the decisive role. However, the paper by Colla et al. contains two curves that differ only by measuring frequency (the upper curve from Fig.2 and the upper curve from Fig.3 from Ref.[8]), which can be used to test the frequency dependence given by Eq. (20). These curves attest to the absence of frequency dependence of fatigue, which contradicts Eq. (20).